\documentclass[aps,pra,amsmath,amsfonts,amssymb,twocolumn,superscriptaddress,showpacs]{revtex4-1}
\usepackage{amssymb,amsmath,amsfonts}
\usepackage{graphicx,psfrag,color}
\usepackage{dcolumn}
\usepackage{bbm}
\usepackage{mathbbol}
\usepackage{braket}

\newcommand{\up}{\uparrow}
\newcommand{\down}{\downarrow}
\newcommand*{\balancecolsandclearpage}{%
  \close@column@grid
  \clearpage
  \twocolumngrid
}

\begin{document}
\flushbottom

\title{Connecting velocity and entanglement in quantum walks}

\author{Alexandre C. Orthey Jr.}
\affiliation{Departamento de F\'isica, Universidade do Estado de Santa Catarina, 89219-710, Joinville, SC, Brazil}
\affiliation{Departamento de F\'isica, Universidade Federal do Paran\'a, 81531-980, Curitiba, PR, Brazil}

\author{Edgard P. M. Amorim}
\email{edgard.amorim@udesc.br} 
\affiliation{Departamento de F\'isica, Universidade do Estado de Santa Catarina, 89219-710, Joinville, SC, Brazil}

\date{\today}

\begin{abstract}
We investigate the relation between transport properties and entanglement between the internal (spin) and external (position) degrees of freedom in one-dimensional discrete time quantum walks. We obtain closed-form expressions for the long-time position variance and asymptotic entanglement of quantum walks whose time evolution is given by any balanced quantum coin, starting from any initial qubit and position states following $\delta$-like (local) and Gaussian distributions. We find out that the knowledge of the limit velocity of the walker together with the polar angle of the initial qubit provide the asymptotic entanglement for local states, while this velocity with the quantum coin phases give it for highly delocalized states. 

\end{abstract}

\pacs{03.67.Ac, 05.60.Gg, 03.67.Bg}

\maketitle

\section{Introduction} \label{sec.1}

Quantum random walks \cite{aharonov1993quantum} or quantum walks are known as the quantum counterparts of the classical random walks. The walker is a quantum particle with a spin-$1/2$ state (qubit) as an internal degree of freedom placed on a regular lattice where each site corresponds to an external degree of freedom (position). Instead of tossing a coin to determine whether the particle goes to the left or right, the time evolution is given by a unitary operator applied successive times to the initial quantum walk state. This operator is constituted by a quantum coin and a conditional displacement operator. The quantum coin operates over the qubit by putting it on a new superposition of spin states. After that, the conditional displacement operator displaces the up (down) spin state to the right (left) neighbor position \cite{kempe2003quantum}. 

The main difference between classical random walks and quantum walks is the superposition principle, which allows the latter to have unique features: a double peak probability distribution showing a quadratic gain in their position variance over time and the creation of entanglement between the spin and position \cite{venegas2012quantum}. These properties have many implications in basic science and underlying potential technological applications. For instance, early works have demonstrated that they are useful to perform computational tasks as a quantum search engine \cite{shenvi2003quantum,tulsi2008faster}, to make universal quantum computation \cite{childs2009universal,lovett2010universal}, for the understanding of some biological processes such as photosynthesis \cite{engel2007evidence} or human decision making \cite{busemeyer2006quantum}, to foster entanglement protocols \cite{vieira2013dynamically}, for generating Anderson localization \cite{vieira2014entangling} and to test the foundations of quantum mechanics \cite{robens2015ideal}. Moreover, they are versatile enough to be implemented in some experimental platforms \cite{wang2013physical}. 

The initial quantum walk state can be a qubit over one position (local state) \cite{kempe2003quantum,tregenna2003controlling} or spread over many positions following some sort of distribution function (delocalized state) \cite{valcarcel2010tailoring,romanelli2010distribution,zhang2016creating,orthey2017asymptotic}. The spreading and entanglement in quantum walks are very sensitive to their initial conditions and quantum coin. This dependence is reflected in the probability distribution of the state over time. For instance, while one initial state leads to a symmetrical probability distribution evolved by one kind of coin, another state or coin leads to an asymmetrical one \cite{kempe2003quantum,tregenna2003controlling}. The entanglement content is similar: while one initial state allows the quantum walks to reach the maximal entanglement, another leads to the minimal entanglement \cite{carneiro2005entanglement,abal2006quantum,abal2006quantumE,eryiugit2014time,orthey2017asymptotic}. Peculiarly, while a local state has only two qubits which cause the maximal entanglement, a delocalized state has a continuous set of initial qubits given by simple expressions between the polar and azimuthal angles of the initial qubit \cite{orthey2017asymptotic}.

The relation between the initial qubit and the chirality introduced by the coin (coin bias) in quantum walks on their spreading or entanglement was analyzed in previous works \cite{tregenna2003controlling,valcarcel2010tailoring,romanelli2010distribution,zhang2016creating,carneiro2005entanglement,eryiugit2014time}. However, as far as we know, some differences between local and delocalized states still remain uncovered by the literature, such as the influence of the quantum coin phases, which play an important role in delocalized states. Despite the efforts to characterize the transport and entanglement in quantum walks, the main question that arises here is: what kind of information regarding the entanglement is provided by the transport? Therefore, our main purpose is to make an analytic connection between their limit velocity and asymptotic entanglement to enlighten some aspects of the question above. In an attempt to perform this study as broad as possible, we consider quantum walks starting from any qubit, using two kinds of position states (local and Gaussian), and time-evolved by any balanced quantum coin. 

This article is structured as follows. In Sec.~\ref{sec.2}, we review the mathematical formalism of quantum walks. In Secs.~\ref{sec.3} and~\ref{sec.4}, by means of Fourier analysis \cite{ambainis2001one,brun2003quantum} we obtain general expressions for the long-time variance and asymptotic entanglement, respectively, and in Sec.~\ref{sec.5}, a linkage between them is established and discussed. Finally, a brief conclusion with our main results is depicted in Sec. \ref{sec.6}.  

\section{One-dimensional quantum walks} \label{sec.2}

The quantum walk state belongs to $\mathcal{H}=\mathcal{H}_C\otimes\mathcal{H}_P$, where $\mathcal{H}_C$ is the coin space, a complex two-dimensional vector space spanned by the spin states $\left\{\ket{\up}, \ket{\down}\right\}$, and $\mathcal{H}_P$ is the position space, a countably infinite-dimensional vector space spanned by a set of orthonormal vectors $\left\{\ket{j}\right\}$ with $j\in\mathbb{Z}$ being the discrete positions on a lattice. Then, the one-dimensional quantum walker has a qubit (spin-$1/2$-like) or coin state,
\begin{equation}
\ket{\Psi_C}=\cos\left(\dfrac{\alpha}{2}\right)\ket{\up}+e^{i\beta}\sin\left(\dfrac{\alpha}{2}\right)\ket{\down},
\label{qubit}
\end{equation}
as the internal degree of freedom in the Bloch sphere representation~\cite{nielsen2010quantum} with polar angle $\alpha\in[0,\pi]$ and azimuth angle $\beta\in[0,2\pi]$, and its position and momentum as external degrees of freedom. Let us consider an initial quantum walk state,
\begin{align}
\ket{\Psi(0)}&=\sum_{j=-\infty}^{+\infty}\ket{\Psi_C}\otimes f(j)\ket{j}\nonumber\\
&=\sum_{j=-\infty}^{+\infty}\left[a(j,0)\ket{\up}+b(j,0)\ket{\down}\right]\otimes \ket{j},
\label{Psi0}
\end{align}
where the initial amplitudes $a(j,0)=f(j)\cos(\alpha/2)$ and $b(j,0)=f(j)e^{i\beta}\sin(\alpha/2)$ correspond to the spins up and down, respectively, $|f(j)|^2$ gives us an initial position distribution function, and $\sum_j[|a(j,0)|^2+|b(j,0)|^2]=1$ over all integers is the condition of normalization.

The dynamical evolution of the quantum walk state is given by $\ket{\Psi(t)}=U(q,\theta,\phi)^t\ket{\Psi(0)}$ in discrete time steps with $U(q,\theta,\phi)=S.[C(q,\theta,\phi)\otimes\mathbbm{1}_P]$ being the time evolution operator where $\mathbbm{1}_P$ is the identity operator in $\mathcal{H}_P$, $C(q,\theta,\phi)$ is the quantum coin, and $S$ is the conditional displacement operator. The quantum coin $C(q,\theta,\phi)$ belongs to the $SU(2)$, and up to an irrelevant global phase, the most general way to write it is \cite{vieira2013dynamically, vieira2014entangling,tregenna2003controlling,carneiro2005entanglement}  
\begin{equation}
\displaystyle
C(q,\theta,\phi) =
\begin{bmatrix}
\sqrt{q}            &  \sqrt{1-q}e^{i\theta} \\
\sqrt{1-q}e^{i\phi} &  -\sqrt{q}e^{i(\theta+\phi)}
\end{bmatrix},
\label{Quantum_coin}
\end{equation}
where the parameters $0\leqslant\theta,\phi\leqslant2\pi$ control the relative phases between spin states and the chirality parameter $0\leqslant q \leqslant1$ determines if the coin is biased ($q\neq 1/2$) or unbiased ($q=1/2$). A fair or unbiased quantum coin $C(\theta,\phi)$ operates over the spin states generating a balanced superposition of them. Two common fair coins are Hadamard and Fourier (Kempe). While the Hadamard coin creates a superposition without relative phases between spin states ($\theta,\phi=0$), the Fourier coin imposes a relative phase of $\pi/2$ ($\theta,\phi=\pi/2$) between them. At last, the conditional displacement operator is
\begin{equation}
S=\sum_j(\ket{\up}\bra{\up}\otimes\ket{j+1}\bra{j}+\ket{\down}\bra{\down}\otimes\ket{j-1}\bra{j}),
\label{S}
\end{equation}
and it shifts the up (down) spin state from site $j$ to site $j+1$ ($j-1$), which generates entanglement between the spin and position states.

The quantum walk state $\ket{\Psi(t)}$ remains pure over time, so the entanglement can be quantified by means of the von Neumann entropy $S_E(\rho(t))=-\mathrm{Tr}[\rho_C(t)\log_2\rho_C(t)]$, where $\rho_C(t)=\mathrm{Tr}_P[\ket{\Psi(t)}\bra{\Psi(t)}]$ is the partially reduced coin state \cite{bennett1996concentrating} and $\mathrm{Tr}_P[\cdot]$ is the trace over the positions; then we have
\begin{equation}
\rho_C(t)=
\begin{bmatrix}
A(t)   & \gamma(t)\\
\gamma^*(t) & B(t)
\end{bmatrix},
\label{rho}
\end{equation}
with $A(t)=\sum_j|a(j,t)|^2$, $\gamma(t)=\sum_ja(j,t)b^*(j,t)$, $\gamma^*(t)$ is the complex conjugate of $\gamma(t)$, and $B(t)=1-A(t)$. The eigenvalues of $\rho_C(t)$ are
\begin{equation}
\Lambda_{\pm}=\dfrac{1}{2} \pm \sqrt{\dfrac{1}{4}-A(t)(1-A(t))+|\gamma(t)|^2};
\label{Lambda}
\end{equation}
therefore,
\begin{equation}
S_E(t)=-\Lambda_+(t)\log_2\Lambda_{+}(t)-\Lambda_{-}(t)\log_2\Lambda_{-}(t),
\label{SE}
\end{equation}
which ranges from zero (separable states) to $1$ (maximal entanglement). 

\section{Long-time variance} \label{sec.3}

After some initial fluctuation, the variance of quantum walks attains the long-time regime. We should make a change of basis to the dual $k$-space $\mathcal{\tilde{H}}_k$ spanned by the Fourier-transformed vectors $\ket{k}=\sum_{j}e^{ikj}\ket{j}$ with $k\in[-\pi,\pi]$ \cite{ambainis2001one} to reach an expression for the long-time variance. The initial state in Eq.~(\ref{Psi0}) is rewritten as
\begin{equation}
\ket{\tilde{\Psi}(0)}=\int_{-\pi}^{\pi} \dfrac{\mathrm{d}k}{2\pi}\ket{\Phi_k(0)}\otimes\ket{k},
\label{Psitil_0}
\end{equation}
where $\ket{\Phi_k(0)}=\tilde{a}_k(0)\ket{\up}+\tilde{b}_k(0)\ket{\down}$. In the Fourier representation, the conditional displacement operator $S$ is diagonal, $S_k=\ket{\up}\bra{\up}\otimes e^{-ik}\ket{k}\bra{k}+\ket{\down}\bra{\down}\otimes e^{ik}\ket{k}\bra{k}$, and thus the time evolution operator in the $k$-space with a fair coin gives
\begin{equation}
U_k=[e^{-ik}\ket{\up}\bra{\up}+e^{ik}\ket{\down}\bra{\down}]C(\theta,\phi).
\label{Uk}
\end{equation}
After diagonalizing $U_k$, to shorten the notation we assume that $k_\delta=k-\delta$, $\delta=(\theta+\phi)/2$ and $\eta=(\theta-\phi)/2$; then we find the following eigenvalues,
\begin{equation}
\lambda_k^{\pm}=\pm\dfrac{e^{i\delta}}{\sqrt{2}}\left[\sqrt{1+\cos^2 k_\delta}\pm i\sin k_\delta\right]=\pm e^{i(\delta \pm \omega)},
\label{eigenvalues}
\end{equation}
since $\sin\omega=\sin k_\delta/\sqrt{2}$ with $\omega\in[-\pi/2,\pi/2]$ and their respective eigenvectors,
\begin{align}
\ket{\Phi_k^{\pm}}&=\dfrac{1}{N_k^{\pm}}
\begin{bmatrix}
e^{ik} \\
e^{-i(\delta+\eta)}\left(\sqrt{2}\lambda_k^{\pm}-e^{ik}\right)
\end{bmatrix},
\label{eigenvectors}
\end{align}
with $(N_k^{\pm})^2=4(1\mp\cos k_\delta\sqrt{1+\cos^2 k_\delta}\pm\sin^2 k_\delta)$. It is possible to expand the states $\ket{\Phi_k(0)}$ in terms of the eigenstates of $U_k$,
\begin{equation}
\ket{\Phi_k(0)}=c_k^+\ket{\Phi_k^+}+c_k^-\ket{\Phi_k^-},
\label{phi_k0+-}
\end{equation}
in such a way that $c_k^{\pm}=\braket{\Phi_k^{\pm}|\Phi_k(0)}$. After following the development from Ref. \cite{brun2003quantum}, we have
\begin{align}
\braket{\hat{\mathbf{j}}^m}_t=&\int_{-\pi}^{\pi} \dfrac{\mathrm{d}k}{2\pi}\left\{ |c_k^+|^2\bra{\Phi_k^+}\hat{Z}\ket{\Phi_k^+}^m\right.\nonumber\\
&+\left. |c_k^-|^2\bra{\Phi_k^-}\hat{Z}\ket{\Phi_k^-}^m \right\}t^m,
\label{jm}
\end{align}
where $\hat{Z}=\ket{\up}\bra{\up}-\ket{\down}\bra{\down}$ and $t\gg 1$, since oscillatory terms are disregarded. The expected values of $\hat{Z}$ are
\begin{equation}
\braket{\Phi_k^{\pm}|\hat{Z}|\Phi_k^{\pm}}=\pm\dfrac{\cos (k-\delta)}{\sqrt{1+\cos^2(k-\delta)}},
\label{expected_value}
\end{equation}
and also the coefficients,
\begin{equation}
c_k^{\pm}=\dfrac{e^{-ik}}{N_k^{\pm}}\{\tilde{a}_k(0)-\tilde{b}_k(0)e^{i(\delta+\eta)}[  1-\sqrt{2}e^{ik}/\lambda_k^{\pm}]\},
\label{ck}
\end{equation}
where the spin-up and -down initial amplitudes are $\tilde{a}_k(0)=\tilde{f}(k)\cos(\alpha/2)$ and $\tilde{b}_k(0)=\tilde{f}(k)e^{i\beta}\sin(\alpha/2)$, respectively, in the $k$-space. By inserting these amplitudes in Eq.~(\ref{ck}) and replacing it together with Eq.~(\ref{expected_value}) in Eq.~(\ref{jm}), for $m=1$ we find the expected position, 
\begin{align}
\braket{\hat{\mathbf{j}}}_t=&\{I(\tilde{f},\delta)[\cos\alpha+\sin\alpha\cos(\beta+\delta+\eta)]\nonumber\\
&-R(\tilde{f},\delta)\sin\alpha\sin(\beta+\delta+\eta)\}t,
\label{j_general}
\end{align}
which depends on the initial qubit given by $\alpha$ and $\beta$ and coin parameters $\delta+\eta$. Similarly, for $m=2$ we have the expected squared position,
\begin{equation}
\braket{\hat{\mathbf{j}}^2}_t=I(\tilde{f},\delta)t^2,
\label{j2_general}
\end{equation}
where the remaining integrals $I(\tilde{f},\delta)$ and $R(\tilde{f},\delta)$ are 
\begin{align}
I(\tilde{f},\delta)&=\int_{-\pi}^{\pi} \dfrac{\mathrm{d}k}{2\pi}|\tilde{f}(k)|^2\left[\dfrac{\cos^2 (k-\delta)}{1+\cos^2 (k-\delta)}\right],
\label{IntegralI}\\
R(\tilde{f},\delta)&=\int_{-\pi}^{\pi} \dfrac{\mathrm{d}k}{2\pi}|\tilde{f}(k)|^2\left[\dfrac{\cos (k-\delta)\sin(k-\delta)}{1+\cos^2 (k-\delta)}\right]. 
\label{IntegralR}
\end{align}

Let us first consider a local state whose initial probability distribution is a Dirac delta function, $|f(j)|^2=\delta(j)$ in Eq.~(\ref{Psi0}), which results $\ket{\Psi_L(0)}=\ket{\Psi_C}\otimes\ket{0}$. The local amplitudes in the $k-$space have $\tilde{f}_L(k)=1$ and inserting it in Eqs.~(\ref{IntegralI}) and~(\ref{IntegralR}) gives $I_L=1-\sqrt{2}/2$ and $R_L=0$, respectively. Then replacing it in Eqs.~(\ref{j_general}) and~(\ref{j2_general}), we can find the long-time variance as 
\begin{align}
&\sigma_L^2(t)=\braket{\hat{\mathbf{j}}^2}_t-\braket{\hat{\mathbf{j}}}_t^2=\left\{\left(1-\dfrac{\sqrt{2}}{2}\right)\right.\nonumber\\
&-\left.\left(\dfrac{3}{2}-\sqrt{2}\right)\left[\cos\alpha+\sin\alpha\cos(\beta+\delta+\eta)\right]^2\right\}t^2
\label{var_local}
\end{align}
for quantum walks starting from an initial local state, arbitrary qubit, and fair coin. The average variance by integrating all qubits over the Bloch sphere results in 
\begin{equation}
\braket{\sigma_L^2}(t)=\int_{0}^{\pi}\dfrac{\mathrm{d}\alpha}{\pi}\int_{-\pi}^{\pi} \dfrac{\mathrm{d}\beta}{2\pi}\sigma_L^2(t)=\dfrac{2\sqrt{2}-1}{8}t^2,
\label{variance_mean_Local}
\end{equation}
and the dependence on the coin parameters vanishes, in agreement with previous works \cite{konno2002quantum,grimett2004weak}.

\begin{table}[b]
\caption{Fitting parameters of $\xi_n$.}
\label{tab.1}
\begin{ruledtabular}
\begin{tabular}{lccccc} 
\hline
            &  $\xi_1$  &  $\xi_2$  &  $\xi_3$  &  $\xi_4$  &  $\xi_5$  \\
\hline 
$\mu$       &  0.8674   & -1.2113   &  0.2477   & -0.6081   &  2.3145   \\     
$\nu$       & -0.6461   &  0.7183   & -0.1083   &  0.4476   & -1.4515   \\
\hline
\end{tabular}
\end{ruledtabular}
\end{table}

\begin{figure}
\includegraphics[width=\linewidth]{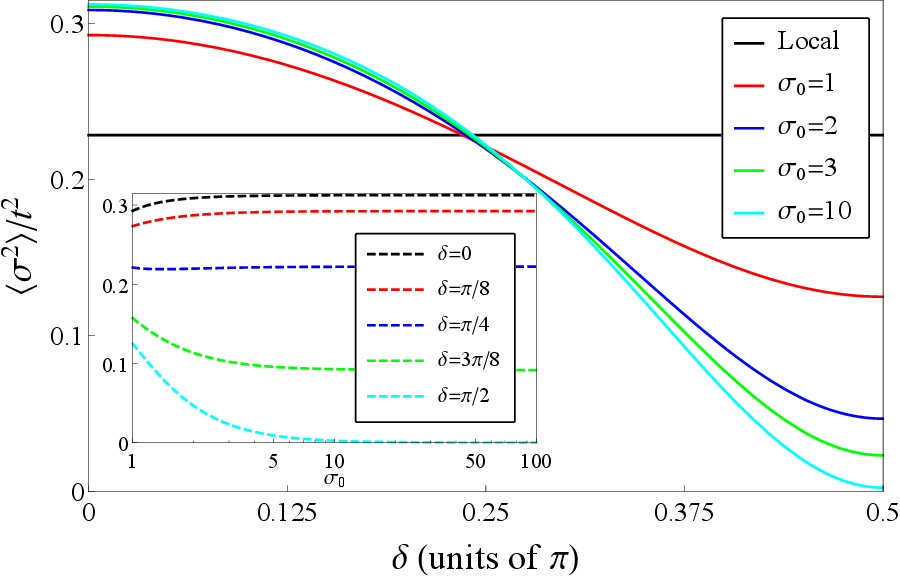}
\caption{The average long-time position variance averaged over all initial qubits for quantum walks starting from a local state (black) from Eq.~(\ref{variance_mean_Local}) and Gaussian states from Eq.~(\ref{variance_mean_Gauss}) with $\sigma_0=1$ (red), $2$ (blue), $3$ (green), and $10$ (cyan) for distinct values of $\delta$. Inset: the average long-time variance of Gaussian states (dashed) for $\delta=0$ (black), $\pi/8$ (red), $\pi/4$ (blue), $3\pi/8$ (green), and $\pi/2$ (cyan) over $\sigma_0$.}
\label{fig.1}
\end{figure}

Let us consider a Gaussian probability distribution with initial dispersion $\sigma_0$. The initial Gaussian state is
\begin{equation}
\ket{\Psi_G(0)}=\sum_{j=-\infty}^{+\infty}\ket{\Psi_C}\otimes\dfrac{e^{-j^2/(4\sigma_0^2)}}{(2\pi\sigma_0^2)^{\frac{1}{4}}}\ket{j},
\label{Psi_0_Gauss}
\end{equation}
where $\sigma_0\geqslant1$; then the initial dispersion is equal to or larger than the distance between two adjacent positions. Since the numerical difference between the discrete summation and integration of the Gaussian amplitudes is around $10^{-4}$ with $\sigma_0=1$, to determine the Gaussian amplitudes in $k-$space, we can change from $j$ to $x$ to integrate
\begin{equation}
\tilde{f}_G(k,\sigma_0)=\int\limits_{-\infty}^{+\infty}\frac{e^{-\left[x^2/(4\sigma_0^2)+ikx \right]}}{\left( 2\pi\sigma_0^2 \right)^{\frac{1}{4}}} \mathrm{d}x=\dfrac{e^{-k^2\sigma_0^2}}{\left(8\pi\sigma_0^2\right)^{-\frac{1}{4}}},
\label{f_Gauss}
\end{equation}
since the imaginary part is gone \cite{orthey2017asymptotic}. After replacing it in Eqs.~\eqref{IntegralI} and \eqref{IntegralR} the remaining integrals do not have exact solutions; however, both approximate numerical solutions bring us
\begin{align}
I_G(\delta,\sigma_0)=&\dfrac{\cos^4\delta}{1+\cos^2\delta}\xi_1(\sigma_0)+\dfrac{\cos^2\delta}{1+\cos^2\delta}[1+\xi_2(\sigma_0)]\nonumber\\
&+\dfrac{\xi_3(\sigma_0)}{1+\cos^2\delta},\label{I_Gauss}\\
R_G(\delta,\sigma_0)=&\dfrac{\cos\delta\sin\delta}{1+\cos^2\delta}[\xi_4(\sigma_0)-1]+\sin(2\delta)\xi_5(\sigma_0),
\end{align}
where $\xi_n(\sigma_0)$ can be fitted by $\mu/\sigma_0^2+\nu/\sigma_0^3$, whose corresponding parameters $\mu$ and $\nu$ are in Table~\ref{tab.1}. Following the same procedure for the local state, by averaging over all qubits, we have
\begin{equation}
\braket{\sigma_G^2}(t)=\left[I_G(\delta,\sigma_0)-\dfrac{3I_G^2(\delta,\sigma_0)+R_G^2(\delta,\sigma_0)}{4}\right]t^2,
\label{variance_mean_Gauss}
\end{equation} 
with a dependence on the quantum coin parameter $\delta$. Neglecting lower-order terms of $\sigma_0^{-3}$, for a Hadamard walk $\braket{\sigma_G^2}\approx[5+2\sum_{n=1}^3\xi_n(\sigma_0)]t^2/16$ with $\braket{\sigma_G^2}\rightarrow 5t^2/16$ for $\sigma_0\gg1$, while for a Fourier walk $\braket{\sigma_G^2}\approx\xi_3(\sigma_0)t^2$ with null variance for large $\sigma_0$. Moreover, for $\sigma_0\rightarrow 0$, $\braket{\sigma_G^2}$ does not converge to $\braket{\sigma_L^2}$ in Eq.~(\ref{variance_mean_Local}), since making it would imply the renormalization of the state by means of a typical error function \cite{orthey2017asymptotic}. However, for $\sigma_0\geqslant1$ the condition of normalization is preserved through this model. Figure~\ref{fig.1} shows $\braket{\sigma^2}/t^2$ for distinct values of $\delta$ given by Eqs.~(\ref{variance_mean_Local}) and~(\ref{variance_mean_Gauss}).

\section{Asymptotic entanglement} \label{sec.4}

\begin{figure}
\includegraphics[width=\linewidth]{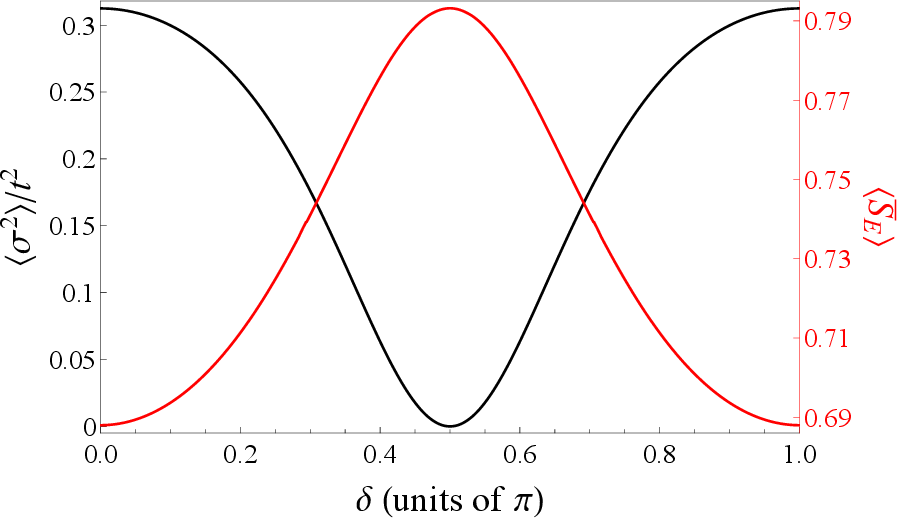}
\caption{The average long-time position variance (black) from Eq.~(\ref{variance_mean_Gauss}) and asymptotic entanglement (red) obtained from Eq.~(\ref{Delta_Gauss}) in $\braket{\overline{S}_E}$ averaged over all initial qubits for quantum walks starting from a Gaussian state with $\sigma_0\gg1$.}
\label{fig.2}
\end{figure}

The entanglement in quantum walks has a strong dependence on the initial conditions and the quantum coin. The asymptotic entanglement, $\overline{S}_E$, which means the asymptotic limit of $S_E(t)$ as $t\rightarrow\infty$, can be evaluated by Fourier analysis. To achieve a general expression, let us first consider the time evolution given by $\ket{\Phi_k(t)}=(U_k)^t\ket{\Phi_k(0)}$ and its spectral decomposition \cite{ambainis2001one}, 
\begin{equation}
\ket{\Phi_k(t)}=e^{i(\delta+\omega)t}c_k^{+}\ket{\Phi_k^{+}}+(-1)^te^{i(\delta-\omega)t}c_k^{-}\ket{\Phi_k^{-}}.
\label{Phikt}
\end{equation}
Therefore, since $\ket{\Phi_k(t)}=(\tilde{a}_k(t),\tilde{b}_k(t))^T$ and the elements of the partially reduced coin state from Eq.~(\ref{rho}) in the $k$-space are 
\begin{equation}
A(t)=\int_{-\pi}^{\pi}\dfrac{\mathrm{d}k}{2\pi}|\tilde{a}_k(t)|^2,
\label{A_t}
\end{equation}
\begin{equation}
\gamma(t)=\int_{-\pi}^{\pi}\dfrac{\mathrm{d}k}{2\pi}\tilde{a}_k(t)\tilde{b}_k^*(t),
\label{gamma_t}
\end{equation}
after replacing Eqs.~(\ref{eigenvectors}) and~(\ref{ck}) into Eq.~(\ref{Phikt}) and by taking $t\rightarrow\infty$, the time dependence vanishes in both Eqs.~(\ref{A_t}) and~(\ref{gamma_t}) giving us $\overline{A}$ and $\overline{\gamma}$. Inserting them in Eq.~(\ref{SE}), we can get $\overline{S}_E(\Delta)$ with $\overline{\Lambda}_{\pm}=(1\pm\sqrt{ \Delta})/2$, a general expression for the asymptotic entanglement as a function of 
\begin{equation}
\Delta=1-4[\overline{A}(1-\overline{A})-\left|\overline{\gamma}\right|^2],
\label{Delta}
\end{equation}
which contains all information about the initial state and the quantum coin. For quantum walks starting from a local state, 
\begin{equation}
 \Delta_L(\alpha,\beta,\delta,\eta)=(3-2\sqrt{2})[1+\sin(2\alpha)\cos(\beta+\delta+\eta)],   
\label{Delta_Local}
\end{equation}
and the corresponding $\overline{S}_E$ ranges from $\sim0.736$ to $1$. The values $(\alpha,\beta)=(3\pi/4,0)$ and $(\pi/4,\pi)$ for a Hadamard walk and $(3\pi/4,-\pi/2)$ and $(\pi/4,\pi/2)$ for a Fourier walk imply $ \Delta_L=0$ and, consequently, $\overline{S}_E=1$~\cite{orthey2017asymptotic}. The integration of $\overline{S}_E$ over the Bloch sphere results in $\braket{\overline{S}_E}\sim0.871$ for any quantum coin. For quantum walks starting from a Gaussian state with $\sigma_0\gg1$, 
\begin{equation}
\Delta_G(\alpha,\beta,\delta, \eta)=\dfrac{\left[\cos\alpha\cos\delta+\sin\alpha\cos(\beta+\eta)\right]^2}{1+\cos^2\delta},
\label{Delta_Gauss}
\end{equation}
which depends on $\delta$ and $\eta$. In this case, the $\overline{S}_E$ ranges from zero up to $1$ and the maximum entanglement condition is given by a continuous set of initial qubits, such that $\cos\beta=-\cot\alpha$ for a Hadamard walk and $\beta=\pm\pi/2$ for any $\alpha$ for a Fourier walk~\cite{orthey2017asymptotic}. After integrating, we have $\braket{\overline{S}_E}\sim0.688$ for a Hadamard walk and $\braket{\overline{S}_E}\sim0.793$ for a Fourier walk being the minimum and maximum values, respectively, as shown in Fig.~\ref{fig.2}, which also displays the average long-time variance $\braket{\sigma^2}/t^2$ for comparison.

\section{Discussion} \label{sec.5}

\begin{figure*}
\includegraphics[width=0.8\linewidth]{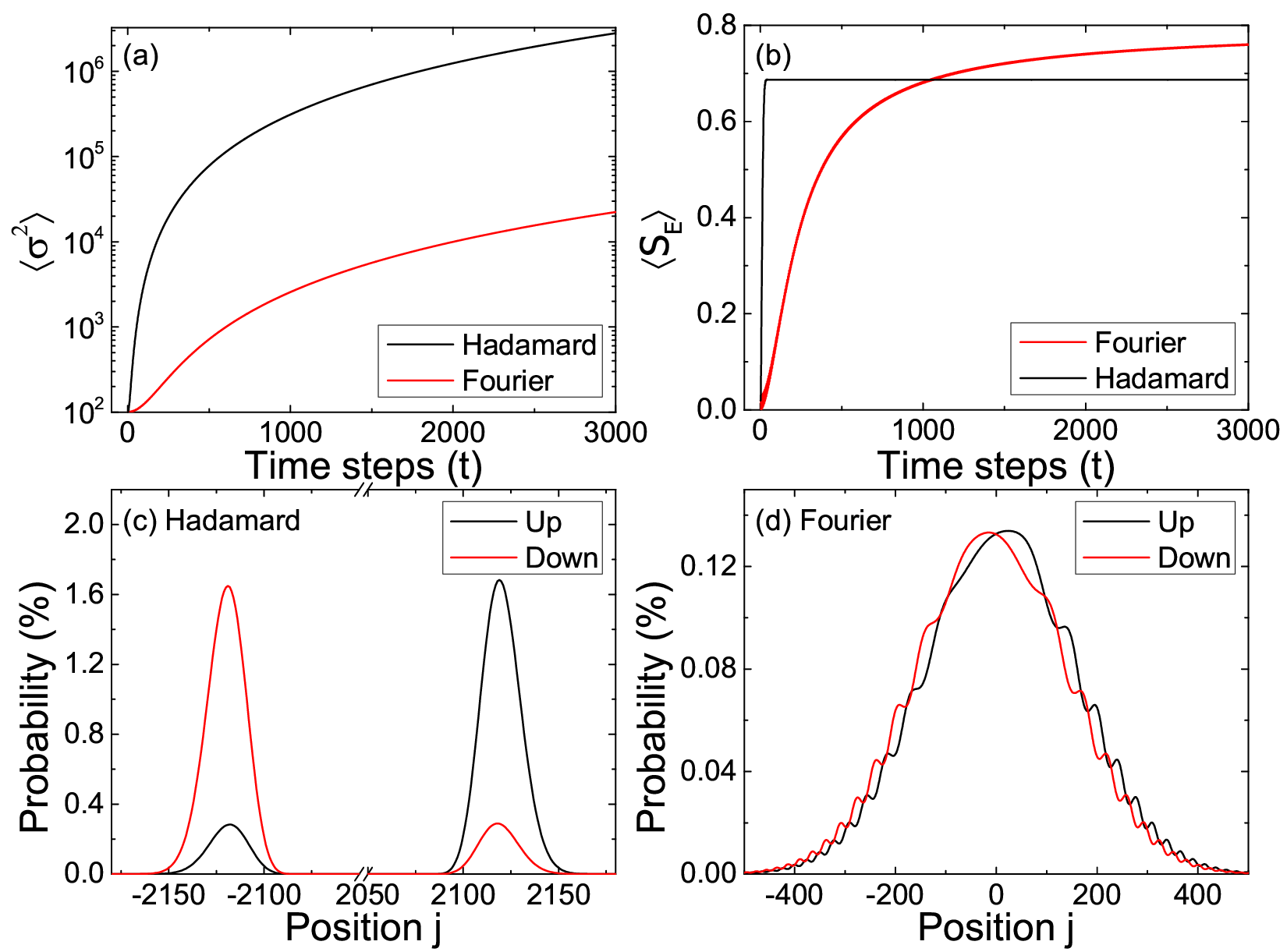}
\caption{Average (a) variance and (b) entanglement over time for Hadamard (black) and Fourier (red) quantum walks starting from a Gaussian state with $\sigma_0=10$. Probability amplitudes for up (black) and down (red) for (c) Hadamard and (d) Fourier quantum walks after $N=3000$ time steps. There is a break region within $j=-2050$ and $2050$ in (c). The average was made over a set of $2016$ initial qubits [Eq.~(\ref{qubit})] from $(\alpha,\beta)=(0,0)$ to $(\pi,2\pi)$ with independent increments of $0.1$.}
\label{fig.3}
\end{figure*}

The average results presented here show some curious differences between quantum walks starting from local and Gaussian states. The average long-time variance and asymptotic entanglement have no dependence on the quantum coin for the quantum walks starting from a local state, but they have strong dependence on the coin parameter $\delta$ when they start from Gaussian states, as shown in Figs.~\ref{fig.1} and~\ref{fig.2}. For instance, the average position variance from a local state is lower than the one from Gaussian states driven by a Hadamard coin ($\delta=0$), while the variance is null by means of a Fourier coin ($\delta=\pi/2$) being stationary for a large initial dispersion, as displayed in Fig.~\ref{fig.2}. In fact, this contrasting behavior also appears in the average entanglement, since Hadamard and Fourier walks have the lowest and highest average entanglement, respectively, opposite to the spreading behavior. Therefore, it is reasonable to state that these two sorts of walks represent borderline spreading and entanglement behaviors.

We carry out extensive numerical calculations of average entanglement and variance over time of Hadamard and Fourier walks starting from a Gaussian state with $\sigma_0=10$, as performed in earlier works \cite{vieira2013dynamically,vieira2014entangling,orthey2017asymptotic,ghizoni2019trojan}. The average variance and entanglement over time depicted in Figs.~\ref{fig.3}(a) and (b), respectively, corroborate the long-time variance and asymptotic entanglement from the previous section. The average variances $\braket{\sigma^2}/t^2$ are about $0.31$ and $<0.01$, while the average entanglements $\braket{S_E}$ reached after $3000$ time steps are about $0.69$ and $0.76$ for Hadamard and Fourier walks, respectively. The probability profile has two opposite peaks in the Hadamard walk as shown in Fig.~\ref{fig.3}(c) with up- and down-state contributions. When the peaks are separated, the entanglement quickly reaches a steady behavior with a large real coherence term from eq. \ref{rho}, since the Hadamard coin does not impose a relative phase between spin states. In the Fourier case, due to the tiny spreading of the state, there is a high overlapping between up and down states as can be seen in Fig.~\ref{fig.3}(d). This interesting effect causes a slow entanglement convergence, and together with the relative phase of $\pi/2$ imposed by the Fourier coin, the real and imaginary parts of the coherence term evolve to a smaller $|\gamma|^2$ than the Hadamard walk.

The expected position in the long-time limit $\braket{\hat{\mathbf{j}}}_t$ in Eq.~(\ref{jm}) is a function of the operator $\hat{Z}$, so it has a dependence on the main diagonal of the partially reduced coin state given by Eq.~(\ref{rho}). Then, by using Eq.~(\ref{A_t}) with $A(t)=1-B(t)$ and after taking the asymptotic limit, we can find $\braket{\hat{\mathbf{j}}}_t=(2\overline{A}-1)t$. Replacing this result in Eq.~(\ref{Delta}), we get $\Delta=(\braket{\hat{\mathbf{j}}}/t)^2+4|\overline{\gamma}|^2$, therefore, since the maximal entanglement in the asymptotic limit $\overline{S}_E\rightarrow1$ would imply $\overline{\gamma}\rightarrow0$ and $\overline{A}\rightarrow1/2$. This result reveals that $\braket{\hat{\mathbf{j}}}_t=0$ is a necessary but not sufficient condition to achieve maximal entanglement. On the other hand, the conditions that lead $\braket{\hat{\mathbf{j}}}_t$ far from the origin position have worse asymptotic entanglement. Let us define the slope of the expected position, $u=\mathrm{d}\braket{\hat{\mathbf{j}}}_t/\mathrm{d}t$, as the limit velocity of the walker \cite{konno2002quantum,grimett2004weak},
\begin{equation}
u_L=\left(1-\dfrac{\sqrt{2}}{2}\right)\left[\cos\alpha+\sin\alpha\cos(\beta+\delta+\eta)\right],
\label{u_local}
\end{equation}
for quantum walks starting from a local state, and also, 
\begin{equation}
u_G=\dfrac{\cos\alpha\cos^2\delta+\sin\alpha\cos\delta\cos(\beta+\eta)}{1+\cos^2\delta},
\label{u_Gauss}
\end{equation}
from a Gaussian state with $\sigma_0\gg1$. After replacing $u_L$ from Eq.~\eqref{u_local} in Eq.~\eqref{Delta_Local}, we get
\begin{equation}
 \Delta_L(\alpha,u_L)=(4-2\sqrt{2})u_L\cos\alpha
-(3-2\sqrt{2})\cos(2\alpha),
\label{Deltau_Local}
\end{equation}
which implies that all dependence on the coin and the relative phase $\beta$ is inside $u_L$. All qubits are within two $\overline{S}_E$ curves with negative concavities given by the conditions $\Delta_L=2u^2_L$ (upper bound) and $\Delta_L=u^2_L+(3-2\sqrt{2})$ (lower bound). In the case of quantum walks starting from highly delocalized Gaussian states, $\Delta_G=2u^2_G$ for a Hadamard walk and for $\delta\rightarrow\pi/2$, so $u_G\rightarrow0$ for a Fourier walk. After replacing $u_G$ from Eq.~\eqref{u_Gauss} in Eq.~\eqref{Delta_Gauss}, we have 
\begin{equation}
 \Delta_G(\delta,u_G)=\dfrac{1+\cos^2\delta}{\cos^2\delta}u^2_G\quad\text{for}\quad\delta\neq\pi/2,
\label{Deltau_Gauss}
\end{equation}
and it means that the maximal entanglement condition is achieved whenever the expected position is null or for $\sin\alpha\cos(\beta+\eta)=0$ with $\delta=\pi/2$. In other words, except for $\delta=\pi/2$, a symmetrical probability distribution which has $\braket{\mathbf{\hat{j}}}=0$, and thus the limit velocity $u_G=0$, is necessary and sufficient to reach the maximal entanglement between the spin and position for highly delocalized Gaussian states. 

\begin{figure}
\includegraphics[width=\linewidth]{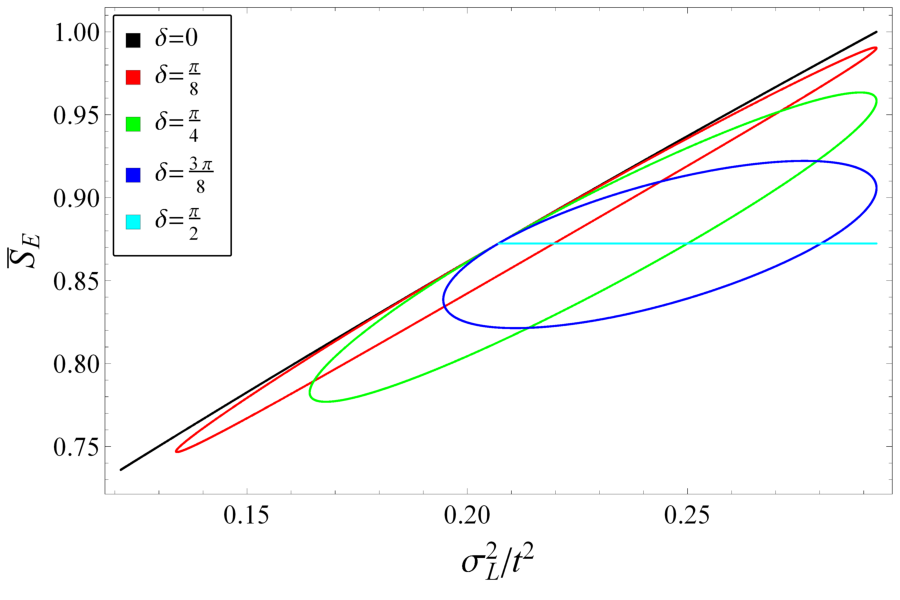}
\caption{Asymptotic entanglement and long-time variance for quantum walks starting from a local state with $\beta=\eta=0$ for $\delta=0$ (black), $\pi/8$ (red), $\pi/4$ (green), $3\pi/8$ (blue), and $\pi/2$ (cyan).}
\label{fig.4}
\end{figure}

\begin{figure}
\centering
\includegraphics[width=\linewidth]{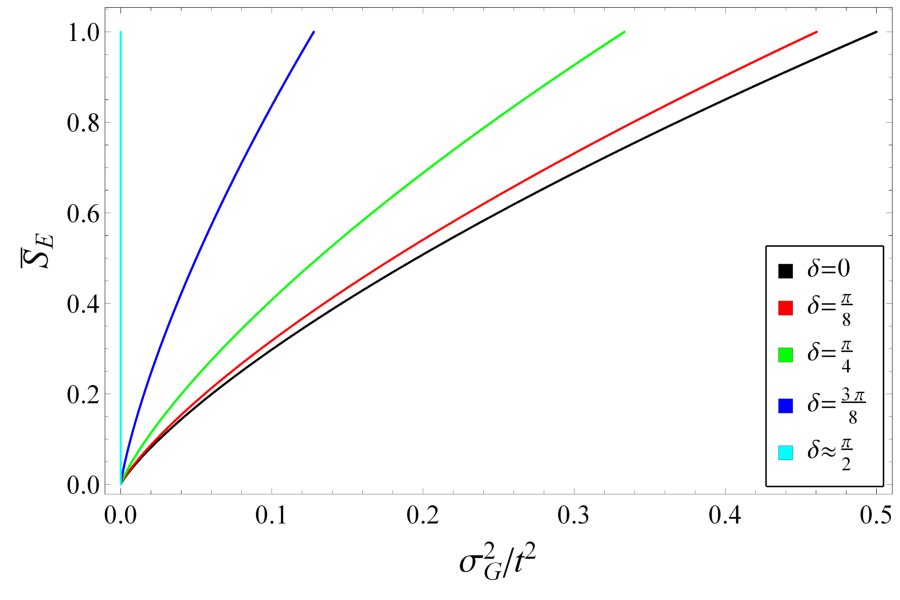}
\caption{Asymptotic entanglement and long-time variance for quantum walks starting from a Gaussian state ($\sigma_0\gg1$) with $\beta=\eta=0$ for $\delta=0$ (black), $\pi/8$ (red), $\pi/4$ (green), $3\pi/8$ (blue), and $0.4995\pi$ (cyan).}
\label{fig.5}
\end{figure}

The expressions as a function of the initial qubit reached so far allow us to correlate the long-time spreading behavior of the states to their respective asymptotic entanglement. Figures~\ref{fig.4} and~\ref{fig.5} show scatter plots between the asymptotic entanglement and the long-time variance, where each point corresponds to a distinct initial qubit given by $\alpha$ from zero to $\pi$ without relative phase between spin states ($\beta=0$). The local case depicted in Fig.~\ref{fig.4} has the intersection point between all curves corresponding to $\alpha=0$. For Hadamard and Fourier walks, the ellipse curves have unitary eccentricity and distinct axes, varying between these two walks in the intermediate cases. While the Hadamard walk reaches a considerable range of asymptotic entanglement values, the Fourier case has $\overline{S}_E\approx0.87$ for all initial qubits, such as the Hadamard walk for $\beta=\pi/2$ \cite{orthey2017asymptotic}. For the Gaussian case in Fig.~\ref{fig.5}, the entanglement is between $\overline{S}_E=0$ and $1$ for all cases with an appreciable decrease in the long-time variance from the Hadamard to the Fourier walk.

It is worth mentioning that a highly delocalized Gaussian state is distinct from a uniform state, because this last one has all position states occupied for all the time steps, which would imply a closed path geometry or the introduction of any sort of boundary condition. In this scenario, quantum walks starting from a local state have dispersion and entanglement periodic or quasiperiodic over time with a strong dependence on the size of the confinement region \cite{tregenna2003controlling,carneiro2005entanglement,rohde2012entanglement}. We can introduce a boundary through the parameter $q$ from the quantum coin, which controls the relative velocities between the two dominant peaks in the probability distribution as $\pm2\sqrt{q}$ \cite{kempf2009group} for local states and Gaussian states with $\delta=0$. The coin at each position $j$ behaves like a scattering center transmitting the spin up to $j+1$ and reflecting the spin down to $j-1$ with the same probability for $q=1/2$. This generates a probability distribution with relative velocities $\pm\sqrt{2}$ between the two peaks. On the one hand, for $q=0$ the reflection is maximal, and this means that the spin state which comes from the left (right) is reflected to the right (left) trapping the state \cite{ghizoni2019trojan, Li2013position}. On the other hand, for $q=1$, the two spin states are split, generating two peaks with relative velocities $\pm2$. In this case, the initial qubit completely determines the probability of each peak centered at $j=\pm t$, where $t$ is the number of time steps. Then the limit velocity $u=\pm1$ and the entanglement is null for an initial spin-up or -down state, and $u=0$ and the entanglement is maximal for an initial equal superposition between spin states. Therefore, the parameter $q$ has a meaningful impact on the transport and entanglement for both kinds of initial position states \cite{tregenna2003controlling,valcarcel2010tailoring,romanelli2010distribution,zhang2016creating,carneiro2005entanglement,eryiugit2014time,ghizoni2019trojan}.

In view of the possibility to measure the entanglement and the expected position, we believe that our findings can be tested on different experimental platforms \cite{wang2013physical}. In particular, it is important to notice that the external degree of freedom could be the $z$ component of orbital angular momentum instead of position $j$. In this context, the experiments based on the manipulation of the orbital angular momentum of photons from a unique light beam \cite{zhang2010implementation,goyal2013implementing,cardano2015quantum} seem to be promising for implementing delocalized states, besides the local ones, and for testing our results here in a straightforward way. The photon polarization can be written as $\rho_C(t)=\mathbb{1}_C+\sum_{l=1}^{3}r_l\sigma_l$ where $\sigma_l$ are the Pauli matrices. By a properly disposition of half- and quarter-wave plates, it is possible to measure the average polarization of the photon in the vertical-horizontal axis ($r_3$), in the $\pm 45^o$ axis ($r_1$), and the average right-left circular polarization ($r_2$) \cite{peres2002quantum}. After that, a postprocessing by tracing out the external degrees of freedom could be performed to obtain $\rho_C(t)$ \cite{vieira2013dynamically}. In the same way, the measurement of the expected position of state, together with the knowledge of the initial qubit and the quantum coin both given by the experimental arrangement, can supply a different route to establish the entanglement.

\section{Conclusion} \label{sec.6}

Quantum walks have a rich spreading and entanglement behavior at long times with a deep dependence on the initial conditions and quantum coin. We have considered quantum walks starting from local and Gaussian states by means of Fourier analysis to find general closed-form expressions for the long-time variance and asymptotic entanglement. From these results, we have pointed out some peculiarities that distinguish local and Gaussian states. First, by averaging over all initial qubits, we showed that the average variance is constant from local states, while from Gaussian states have strong dependence on the quantum coin, being stationary for a Fourier walk with high initial delocalization. Second, since there are two initial qubits from local states and a continuous set of qubits from Gaussian states which lead to the maximal entanglement evolved by means of Hadamard and Fourier coins \cite{orthey2017asymptotic}, we extended these results here for any fair quantum coin. Third, we also corroborated our analytical results through numerical calculations to verify the average spreading and entanglement behavior over time of Hadamard and Fourier walks.

After obtaining the long-time position variance and asymptotic entanglement, we established a linkage between them by detaching the dependency on the initial conditions and quantum coin, and by showing their correlation via scatter plots. Our main result shows that the achievement of the limit velocity of the walker together with the knowledge of the polar angle of the initial qubit (coin state) furnish the asymptotic entanglement for local states, and the limit velocity with the quantum coin phases provide it for highly delocalized states. Furthermore, we believe that better understanding of the entanglement content from the quantum transport behavior perspective might offer novel ways to build entanglement protocols and to interpret some measurements.

\begin{acknowledgments}
This study was financed in part by the Coordena\c{c}\~ao de Aperfei\c{c}oamento de Pessoal de N\'ivel Superior - Brasil (CAPES) - Finance Code 001. A. C. O. thanks E. L. Brugnago and R. M. Angelo for insightful suggestions. E. P. M. A. thanks J. Longo for her careful reading and corrections.
\end{acknowledgments}


\begin{thebibliography}{200}

\bibitem{aharonov1993quantum} Y. Aharonov, L. Davidovich, and N. Zagury,
Phys. Rev. A \textbf{48}, 1687 (1993).
\bibitem{kempe2003quantum} J. Kempe, Contemp. Phys. \textbf{44}, 307 (2003).
\bibitem{venegas2012quantum} S. E. Venegas-Andraca, Quantum Inf. Process. \textbf{11}, 1015 (2012).
\bibitem{shenvi2003quantum} N. Shenvi, J. Kempe, and K. Birgitta Whaley, Phys. Rev. A \textbf{67}, 052307 (2003).
\bibitem{tulsi2008faster} A. Tulsi, Phys. Rev. A \textbf{78}, 012310 (2008).
\bibitem{childs2009universal} A. M. Childs, Phys. Rev. Lett. \textbf{102}, 180501 (2009).
\bibitem{lovett2010universal} N. B. Lovett, S. Cooper, M. Everitt, M. Trevers, and V. Kendon, Phys. Rev. A \textbf{81}, 042330 (2010).
\bibitem{engel2007evidence} G. S. Engel, T. Calhoun, E. L. Read, T.-K. Ahn, T. Man{\v{c}}al, Y.-C. Cheng,  R. E. Blankenship, and G. R. Fleming, Nature \textbf{446}, 782, (2007).
\bibitem{busemeyer2006quantum} J. R. Busemeyer, Z. Wang, and J. T. Townsend, J. of Math. Psych. \textbf{50}, 220 (2006).
\bibitem{vieira2013dynamically} R. Vieira, E. P. M. Amorim, and G. Rigolin, Phys. Rev. Lett. \textbf{111}, 180503, (2013).
\bibitem{vieira2014entangling} R. Vieira, E. P. M. Amorim, and G. Rigolin, Phys. Rev. A \textbf{89}, 042307, (2014).
\bibitem{robens2015ideal} C. Robens, W. Alt, D. Meschede, C. Emary, and A. Alberti, Phys. Rev. X \textbf{5}, 011003 (2015).
\bibitem{wang2013physical} J. Wang and K. Manouchehri, \textit{Physical implementation of quantum walks} (Springer, 2013).
\bibitem{tregenna2003controlling} B. Tregenna, W. Flanagan, R. Maile, and V. Kendon, New J. of Phys. \textbf{5}, 83 (2003).
\bibitem{valcarcel2010tailoring} G. J. de Valc{\'a}rcel, E. Rold{\'a}n, and A. Romanelli, New J. of Phys. \textbf{12}, 123022 (2010).
\bibitem{romanelli2010distribution} A. Romanelli, Phys. Rev. A \textbf{81}, 062349 (2010).
\bibitem{zhang2016creating} W.-W. Zhang, S. K. Goyal, F. Gao, B. C. Sanders, and C. Simon, New J. of Phys. \textbf{18}, 093025 (2016).
\bibitem{orthey2017asymptotic} A. C. Orthey and E. P. M. Amorim, Quantum Inf. Process. \textbf{16}, 224 (2017).
\bibitem{carneiro2005entanglement} I. Carneiro, M. Loo, X. Xu, M. Girerd, V. Kendon, and P. L. Knight, New J. Phys. \textbf{7}, 156 (2005).
\bibitem{abal2006quantum} G. Abal, R. Siri, A. Romanelli, and R. Donangelo, Phys. Rev. A \textbf{73}, 042302 (2006).
\bibitem{abal2006quantumE}  G. Abal, R. Siri, A. Romanelli, and R. Donangelo, Phys. Rev. A \textbf{73}, 069905(E) (2006).
\bibitem{eryiugit2014time} R. Eryi{\u{g}}it and S. G{\"u}nd{\"u}{\c{c}}, Int. J. of Quantum Inf. \textbf{12}, 1450036 (2014).
\bibitem{ambainis2001one} A. Ambainis, E. Bach, A. Nayak, A. Vishwanath, and J. Watrous, in \textit{Proceedings of the thirty-third annual ACM symposium on Theory of computing} (ACM, 2001) pp. 37-49.
\bibitem{brun2003quantum} T. A. Brun, H. A. Carteret, and A. Ambainis, Phys. Rev. Lett. \textbf{91}, 130602 (2003).
\bibitem{nielsen2010quantum} M. A. Nielsen and I. L. Chuang, \textit{Quantum computation and quantum information, 10th Anniversary edition} (Cambridge University Press, Cambridge, 2010).
\bibitem{bennett1996concentrating} C. H. Bennett, H. J. Bernstein, S. Popescu, and B. Schumacher, Phys. Rev. A \textbf{53}, 2046 (1996).
\bibitem{konno2002quantum} N. Konno, Quantum Inf. Process. \textbf{1}, 345 (2002).
\bibitem{grimett2004weak} G. Grimmett, S. Janson, and P. F. Scudo, Phys. Rev. E \textbf{69}, 026119 (2004).
\bibitem{ghizoni2019trojan} H. S. Ghizoni and E. P. M. Amorim, Braz. J. Phys. 49, 168 (2019).
\bibitem{rohde2012entanglement} P. P. Rohde, A. Fedrizzi, and T. C. Ralph, J. of Mod. Opt. \textbf{59}, 710 (2012).
\bibitem{kempf2009group} A. Kempf and R. Portugal, Phys. Rev. A \textbf{79}, 052317 (2009).
\bibitem{Li2013position} Z. J. Li, J. A. Izaac, and J. B. Wang, Phys. Rev. A, \textbf{87}, 012314 (2013).
\bibitem{zhang2010implementation} P. Zhang, B.-H. Liu, R.-F. Liu, H.-R. Li, F.-L. Li, and G.-C. Guo, Phys. Rev. A, \textbf{81}, 052322 (2010).
\bibitem{goyal2013implementing} S. K. Goyal, F. S. Roux, A. Forbes, and T. Konrad, Phys. Rev. Lett. \textbf{110}, 263602 (2013).
\bibitem{cardano2015quantum} F. Cardano \textit{et al.}, Sci. Adv. \textbf{1}, e1500087 (2015).
\bibitem{peres2002quantum} A. Peres, \textit{Quantum theory: concepts and methods} (Kluwer Academic Publishers, 2002).

\end{thebibliography}
\end{document}